\documentclass[twocolumn]{aastex63}
\usepackage{color}
\usepackage[titletoc]{appendix}
\usepackage[fleqn]{amsmath}
\usepackage{amssymb}
\usepackage{mathtools}
\usepackage{upgreek}
\usepackage{float}
\usepackage{comment}
\usepackage{enumitem}
\usepackage{natbib}
\usepackage{graphicx}
\usepackage{bm}
\usepackage{totcount}
\usepackage{multirow}

\newtotcounter{citnum} 
\def\oldbibitem{} \let\oldbibitem=\bibitem
\def\bibitem{\stepcounter{citnum}\oldbibitem}

\shortauthors{Jensen \& Millholland}
\shorttitle{The Libration Amplitude Bias}

\begin{document} 

\title{Inferred Properties of Planets in Mean-Motion Resonances are Biased by Measurement Noise}

\author{David Jensen}
\affiliation{Department of Physics, Princeton University, Princeton, NJ 08544, USA}

\author[0000-0003-3130-2282]{Sarah C. Millholland}
\affiliation{Department of Physics, Massachusetts Institute of Technology, Cambridge, MA 02139, USA}
\affiliation{MIT Kavli Institute for Astrophysics and Space Research, Massachusetts Institute of Technology, Cambridge, MA 02139, USA}
\affiliation{Department of Astrophysical Sciences, Princeton University, Princeton, NJ 08544, USA}
\email{sarah.millholland@mit.edu}

\begin{abstract}

Planetary systems with mean-motion resonances (MMRs) hold special value in terms of their dynamical complexity and their capacity to constrain planet formation and migration histories. The key towards making these connections, however, is to have a reliable characterization of the resonant dynamics, especially the so-called ``libration amplitude'', which qualitatively measures how deep the system is into the resonance. In this work, we identify an important complication with the interpretation of libration amplitude estimates from observational data of resonant systems. Specifically, we show that measurement noise causes inferences of the libration amplitude to be systematically biased to larger values, with noisier data yielding a larger bias. We demonstrated this through multiple approaches, including using dynamical fits of synthetic radial velocity data to explore how the the libration amplitude distribution inferred from the posterior parameter distribution varies with the degree of measurement noise. We find that even modest levels of noise still result in a slight bias. The origin of the bias stems from the topology of the resonant phase space and the fact that the available phase space volume increases non-uniformly with increasing libration amplitude. We highlight strategies for mitigating the bias through the usage of particular priors. Our results imply that many known resonant systems are likely deeper in resonance than previously appreciated.
\\
\end{abstract}

\section{Introduction}
\label{sec: Introduction}

Mean-motion resonance (MMR) refers to an orbital configuration in which two bodies have orbital periods that form a ratio of small integers. These resonances were first studied in the context of the Solar System \citep{1976ARA&A..14..215P}, most notably the Galilean satellites around Jupiter, which form a 4:2:1 resonant chain. Most known extrasolar planets do not display resonant relationships \citep{2014ApJ...790..146F}, although they are also not particularly rare either. Resonant chains like the famous TRAPPIST-1 system are intriguing examples \citep{2017Natur.542..456G}. Compared to planets found in transit surveys, MMRs are more common in systems discovered with radial velocities (RVs), which contain predominantly giant planets \citep{2011ApJ...730...93W}. 

Resonant planetary systems are valuable because they provide a relic of the system's formation history. That is, MMRs are established through convergent migration, particularly migration arising from planet-disk interactions while the planets are still embedded in the protoplanetary disk \citep{1980ApJ...241..425G, 2007ApJ...654.1110T}. The current state of a resonant system can thus, in principle, provide some insight into the formation conditions and other details of the migration process \citep[e.g.][]{2015AJ....149..167B, 2016Natur.533..509M, 2017MNRAS.469.4613S, 2017A&A...605A..96D, 2019NatAs...3..424M, 2020AJ....160..106H}.

One of the most important diagnostics of a resonance is the ``libration amplitude'', which is qualitatively related to the energy of the resonance and conveys the proximity of the system to ``exact'' resonance. Specifically, the libration amplitude reflects the range of oscillations of the planetary conjunctions around their equilibria. The libration amplitude is thought to provide constraints on how smooth or turbulent the migration was \citep[e.g.][]{2018ApJ...867...75D, 2021A&A...656A.115H}. Resonances with very small libration amplitudes indicate very smooth and dissipative formation \citep{2020AJ....160..106H}, whereas large libration amplitudes could be a consequence of stochastic forcing \citep[e.g.][]{2008ApJ...683.1117A, 2009A&A...497..595R}. A large libration amplitude could also indicate a history of perturbations by another planet \citep[e.g.][]{2021AJ....161..161D} or overstable librations \citep{2014AJ....147...32G, 2022ApJ...925...38N}. 

Given the value of the libration amplitude as a tracer of various formation processes, it is crucial to be able to obtain accurate estimates of this quantity from observations of resonant systems. Unfortunately, \cite{2018AJ....155..106M} showed that this may be in jeopardy. They found preliminary evidence (as detailed in Section \ref{sec: libration amplitude bias}) that the libration amplitude inferred from data of resonant systems may be systematically biased to larger values due to measurement uncertainties, in a similar way as eccentricity inferences are affected by measurement noise \citep[e.g.][]{1971AJ.....76..544L, 2008ApJ...685..553S, 2010ApJ...725.2166H}. Though this was suggested, it has not yet been confirmed in detail. In this paper, we use multiple approaches of confirming the bias (Sections \ref{sec: exploring the bias} and \ref{sec: synthetic data experiments}) and understanding its origin (Section \ref{sec: discussion}).

\section{The Resonant Libration Amplitude and Bias from Measurement Noise}
\label{sec: libration amplitude bias}

One of the key measures of mean-motion resonance is the ``critical angle'' (also called ``critical argument'' or ``resonant argument''). For two planets in a first-order $p+1:p$ MMR, there are two critical angles,
\begin{equation}
\begin{split}
\phi_{12,1} &= (p+1)\lambda_2 - p\lambda_1 - \varpi_1 \\
\phi_{12,2} &= (p+1)\lambda_2 - p\lambda_1 - \varpi_2,
\end{split}
\end{equation}
where $\lambda_1$ and $\lambda_2$ are the mean longitudes of the inner and outer planets and $\varpi_1$ and $\varpi_2$ are the longitudes of pericenter. The critical angles describe the evolution of the planetary conjunctions with respect to the pericenters of the two orbits. When a system is in resonance, one or more of the critical angles undergo librations (bounded oscillations) about their equilibria. The ``resonant libration amplitude'' is the amplitude of these oscillations and is related to the total energy of the system, with smaller amplitudes corresponding to lower energies \citep{1999ssd..book.....M}. Systems with zero libration amplitude are maximally damped to their resonant fixed points.

\begin{figure}
\centering
\epsscale{1.1}
\plotone{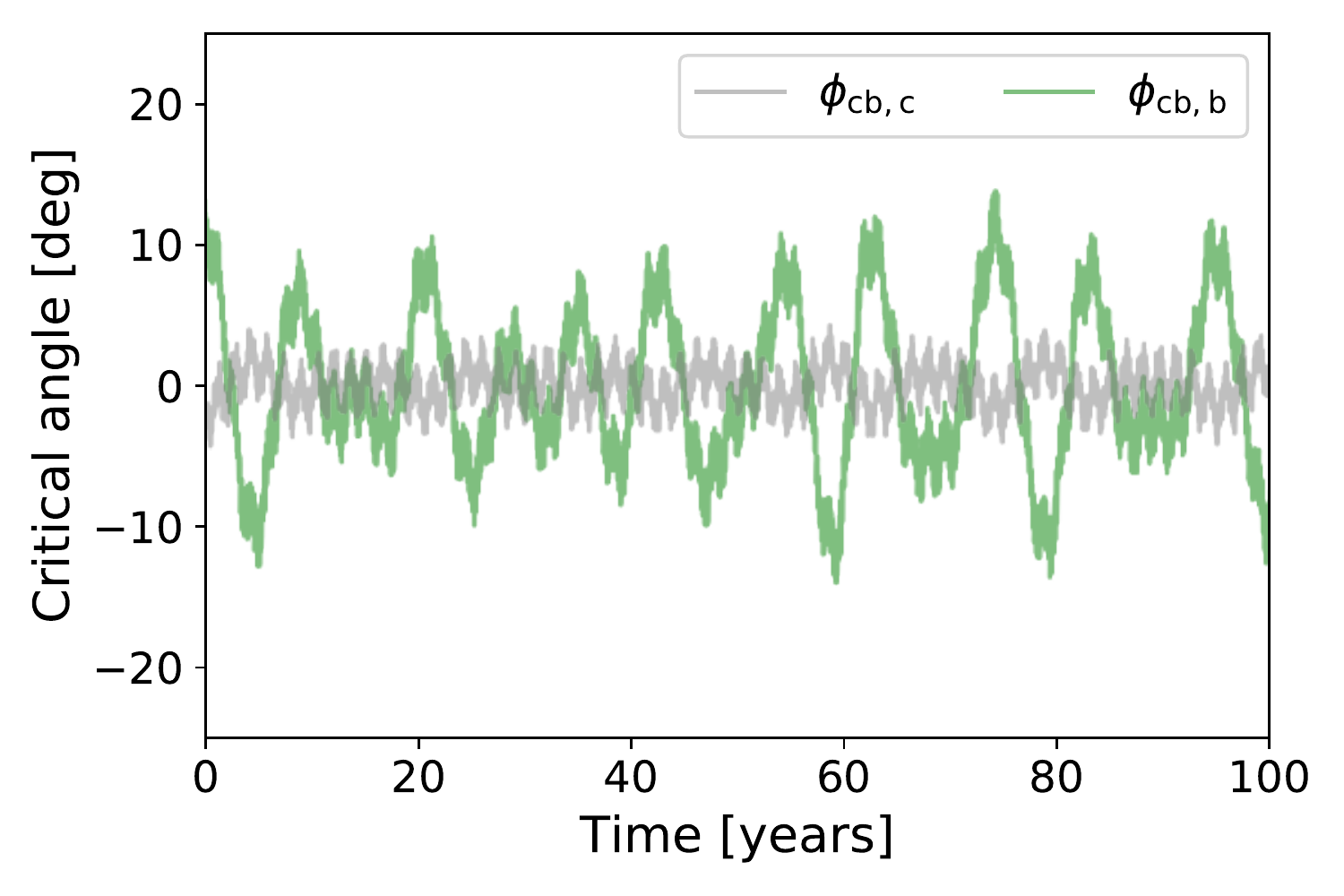}
\caption{Evolution of two critical resonant angles resulting from a numerical integration of the GJ 876 system. The two angles correspond to the 2:1 MMR between planets c and b. 
} 
\label{fig: GJ 876 resonant angles}
\end{figure}

\begin{figure}[t]
\centering
\epsscale{1.1}
\plotone{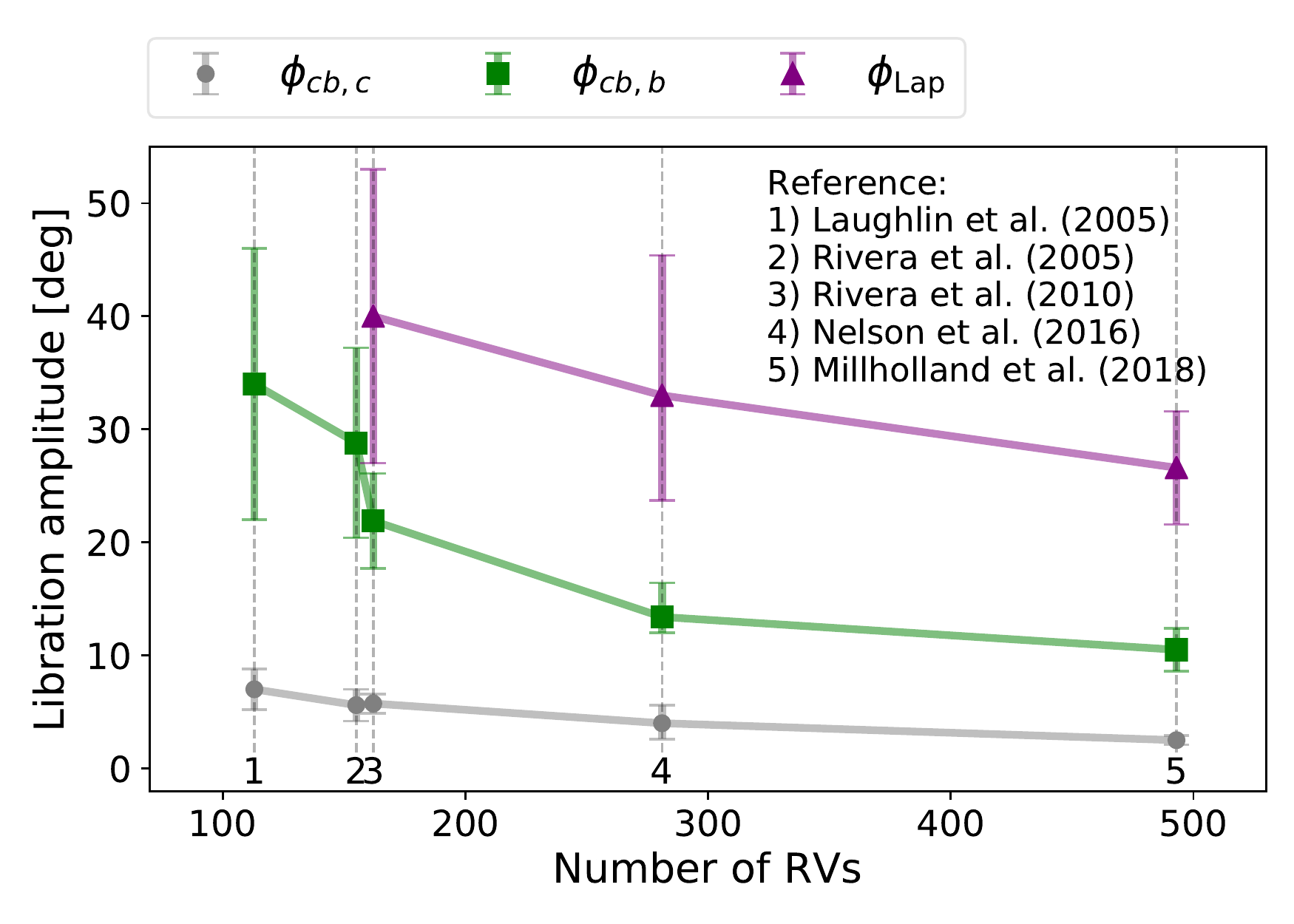}
\caption{Published estimates of the libration amplitudes of three critical resonant angles ($\phi_{cb,c}$, $\phi_{cb,c}$, $\phi_{\mathrm{Lap}}$) in the GJ 876 system, plotted as a function of the number of radial velocity measurements used in the analysis. These results are taken from \cite{2005ApJ...622.1182L}, \cite{2005ApJ...634..625R}, \cite{2010ApJ...719..890R}, \cite{2016MNRAS.455.2484N}, and \cite{2018AJ....155..106M}. Measurements from ELODIE,
CORALIE, or Lick Observatory are not included in the number of RVs. (Figure adapted from \citealt{2018AJ....155..106M}.)} 
\label{fig: libration amplitude vs number of RVs}
\end{figure}

\cite{2018AJ....155..106M} first identified a possible bias of libration amplitude estimates as part of their detailed characterization of the GJ 876 system. GJ 876 is a nearby M4V dwarf hosting four known planets, the outer three of which are locked in a 4:2:1 Laplace resonance \citep{2001ApJ...556..296M, 2005ApJ...622.1182L, 2005ApJ...634..625R, 2010ApJ...719..890R, 2016MNRAS.455.2484N, 2018AJ....155..106M}. The 2:1 MMR of planets c and b (the second and third planets from the star with $P_c \sim 30$ days and $P_b \sim 61$ days) was the first resonance discovered in an exoplanetary system. This MMR has two critical angles,
\begin{equation}
\begin{split}
\phi_{cb,c} &= 2\lambda_b - \lambda_c - \varpi_c \\ 
\phi_{cb,b} &= 2\lambda_b - \lambda_c - \varpi_b.
\end{split}
\end{equation}
In Figure \ref{fig: GJ 876 resonant angles}, we plot the evolution of $\phi_{cb,c}$ and $\phi_{cb,b}$ for a 100 year $N$-body integration of the GJ 876 system using the best-fit parameters identified in \cite{2018AJ....155..106M} as initial conditions. We use the REBOUND gravitational dynamics software package \citep{2012A&A...537A.128R} with the ``WHFast'' Wisdom-Holman symplectic integrator \citep{1991AJ....102.1528W, 2015MNRAS.448L..58R}. The angles undergo low amplitude librations around $0^{\circ}$. There are additional angles analogous to $\phi_{cb,c}$ and $\phi_{cb,b}$ for the 2:1 resonance of planets b and e (the third and fourth planets from the star). Beyond the individual two-body critical angles, the three-body 4:2:1 Laplace resonance of planets c, b, and e is further defined by the libration of the critical angle,
\begin{equation}
\phi_{\mathrm{Lap}} = \lambda_c - 3\lambda_b + 2\lambda_e. 
\end{equation}

Given the long history of explorations of GJ 876 by different research teams, one can explore how the libration amplitude estimates have changed with the size of the RV datasets. \cite{2018AJ....155..106M} showed that the reported libration amplitudes decreased monotonically with each successive characterization of the system. Figure \ref{fig: libration amplitude vs number of RVs} shows the amplitude estimates of $\phi_{cb,c}$, $\phi_{cb,c}$, and $\phi_{\mathrm{Lap}}$ from different publications as a function of the number of RV measurements used in the analyses. Each study deemed the system to be deeper in resonance than all previous studies. Since a larger dataset corresponds to a higher signal-to-noise ratio, this finding may indicate that measurement noise causes resonant systems to appear to have larger libration amplitudes than they actually do. The true state of the system may be even lower energy than the latest measurements indicate.

The above hypothesis -- that measurement noise biases libration amplitude estimates -- needs to be confirmed with further analyses. One reason for this is that the different publications referenced in Figure \ref{fig: GJ 876 resonant angles} used a variety of analysis techniques, including both Bayesian and non-Bayesian methods, so the comparisons between them cannot be directly mapped to differences in signal-to-noise ratios. It would be more instructive to use the same analysis methods on the same dataset and systematically vary either the signal-to-noise ratio or the size of the dataset. Moreover, this would allow us to not only confirm the existence of the bias but also understand its origin. We will explore these concepts in the following sections.

Before proceeding, however, we must clarify exactly what we mean with our usage of the term ``bias''. In frequentist statistics, an estimator $\hat{\theta}$ of a parameter $\theta$ is ``unbiased'' if $\mathrm{Bias}(\theta) = \mathrm{E}(\hat{\theta}) - \theta = 0$, or in other words, if the expected value of the estimator is equal to the true value of the parameter being estimated. The concept of bias is ill-defined in Bayesian statistics, in part because the parameter itself is not considered to be fixed, but rather it is a random variable whose probability distribution we wish to estimate with the inclusion of the prior probability. In this paper we use the term ``bias'' loosely in order to refer to the phenomenon in which progressively larger measurement noise leads the posterior parameter distribution to be increasingly weighted towards larger (or smaller) values. The libration amplitude ``bias'' discussed herein is very analogous to that which plagues the inference of orbital eccentricities \citep[e.g.][]{1971AJ.....76..544L, 2008ApJ...685..553S, 2010ApJ...725.2166H}, although an important difference is that the eccentricity is generally a parameter of the orbit model, whereas the libration amplitude is not.

\section{Exploring the Bias}
\label{sec: exploring the bias}

Although the hypothesized libration amplitude bias was first identified in the GJ 876 system, it is useful to use a simpler resonant system to explore the bias further. For this purpose, we consider the HD 128311 system, which contains two eccentric gas giant planets in a 2:1 MMR \citep{2003ApJ...582..455B, 2005ApJ...632..638V, 2014ApJ...795...41M, 2015MNRAS.448L..58R}. The system was most recently studied by \cite{2015MNRAS.448L..58R}, who performed a dynamical fit and determined the system to be locked in resonance with a libration amplitude of $\sim37^{\circ}$. Some of the relevant best-fit system parameters from \cite{2015MNRAS.448L..58R} are provided in Table \ref{tab: HD 128311 parameters}. In this section, we use \cite{2015MNRAS.448L..58R}'s posterior distributions (H. Rein, private communication) to explore the effects of measurement noise on estimates of the libration amplitude.

In general, lower signal-to-noise data results in broader posterior distributions. Accordingly, we can impose an artificial broadening of the posterior distribution as a means of simulating the effects of added measurement noise without ever touching the raw data. (Later in Section \ref{sec: synthetic data experiments}, we will work directly with the data.) To perform the simulated broadening, we will first demonstrate that the posterior distribution can be well-described by a multivariate Gaussian distribution.  

\begin{table}[t!]
\caption{Best-fit parameters of the HD 128311 system from the dynamical fit by \cite{2015MNRAS.448L..58R}.}
\begin{tabular}{c c}
\hline
\hline
Parameter & Value and $2\sigma$ confidence interval \\ 
\hline
Epoch & 2450983.83 (fixed) \\
$M_{\star}$ & 0.828 $M_{\odot}$ \\
$i$ & ${63.8^{\circ}}^{+23.7^{\circ}}_{-35.9^{\circ}}$ \\ \\
\multicolumn{2}{c}{Planet 1} \\
$M_{p1}\sin i$ & $1.83^{+0.15}_{-0.18} \ M_{\mathrm{Jup}}$ \\
$P_1$ & $460.1^{+4.2}_{-3.6}$ days \\
$e_1$ & $0.30^{+0.03}_{-0.04}$ \\
$\omega_1$ & ${-76.2^{\circ}}^{+6.4^{\circ}}_{-9.2^{\circ}}$ \\
$M_1$ & ${259.2^{\circ}}^{+11.9^{\circ}}_{-12.6^{\circ}}$ \\ \\
\multicolumn{2}{c}{Planet 2} \\
$M_{p2}\sin i$ & $3.20^{+0.08}_{-0.08} \ M_{\mathrm{Jup}}$ \\
$P_2$ & $910.7^{+7.6}_{-6.0}$ days \\
$e_2$ & $0.12^{+0.08}_{-0.06}$ \\
$\omega_2$ & ${-19.7^{\circ}}^{+23.2^{\circ}}_{-12.0^{\circ}}$ \\
$M_2$ & ${184.2^{\circ}}^{+20.0^{\circ}}_{-10.7^{\circ}}$ \\ 
\hline
\label{tab: HD 128311 parameters}
\end{tabular}
\end{table}

\begin{figure*}
\centering
\epsscale{1.1}
\plotone{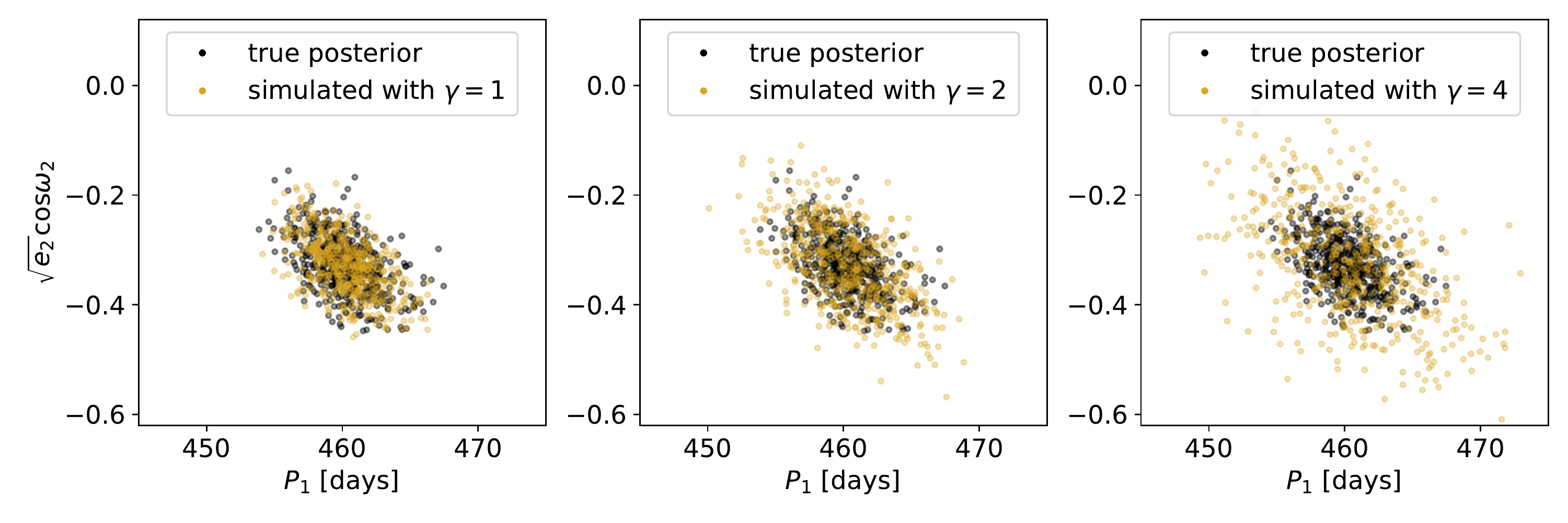}
\caption{One example subplot ($\sqrt{e_2}\cos\omega_2$ vs. $P_1$) of the full corner plot of the posterior distribution, with the three panels indicating three versions of the simulated distribution with different broadening factors $\gamma$. The black (yellow) points correspond to data from the true (simulated) distributions. The simulated distribution in the left panel ($\gamma = 1$, no broadening) agrees well with the true distribution.} 
\label{fig: subplot of corner plot}
\end{figure*}

We parameterize the posterior distribution as
\begin{equation}
\begin{split}
\label{eq: X_post}
\mathbf{X}_{\mathrm{post}} &= 
(\log_{10}M_{p1}\sin i, P_1, \sqrt{e_1}\cos\omega_1, \sqrt{e_1}\sin\omega_1, M_1, \\
&\log_{10}M_{p2}\sin i, P_2, \sqrt{e_2}\cos\omega_2, \sqrt{e_2}\sin\omega_2, M_2, \cos{i}).
\end{split}
\end{equation}
Next, we calculate the mean vector $\boldsymbol{\mu}_{\mathrm{post}}$ and covariance matrix $\boldsymbol{\Sigma}_{\mathrm{post}}$ of $\mathbf{X}_{\mathrm{post}}$ such that the distribution can be closely approximated by a simulated distribution drawn according to the multivariate Gaussian,
\begin{equation}
\mathbf{X}_{\mathrm{sim}} \sim \mathcal{N}(\boldsymbol{\mu}_{\mathrm{post}}, \boldsymbol{\Sigma}_{\mathrm{post}}).
\end{equation}
We use visual inspection of corner plots of the true distribution and a simulated distribution with an equal sample size to verify that the distributions are similar. An example of one sub-plot of this broader corner plot is shown in the left panel of Figure \ref{fig: subplot of corner plot}. 

To approximate the broadening of the posterior distribution that would result from noisier data, we simply scale the covariance matrix by a constant factor, $\gamma > 1$, such that the simulated distribution is now given by 
\begin{equation}
\mathbf{X}_{\mathrm{sim}} \sim \mathcal{N}(\boldsymbol{\mu}_{\mathrm{post}}, \gamma\boldsymbol{\Sigma}_{\mathrm{post}}).
\end{equation}
We explore $\gamma$ values ranging from 1 through 8. Examples of broadened distributions are shown in the middle and right panels of Figure \ref{fig: subplot of corner plot}.

\begin{figure}
\centering
\epsscale{1.1}
\plotone{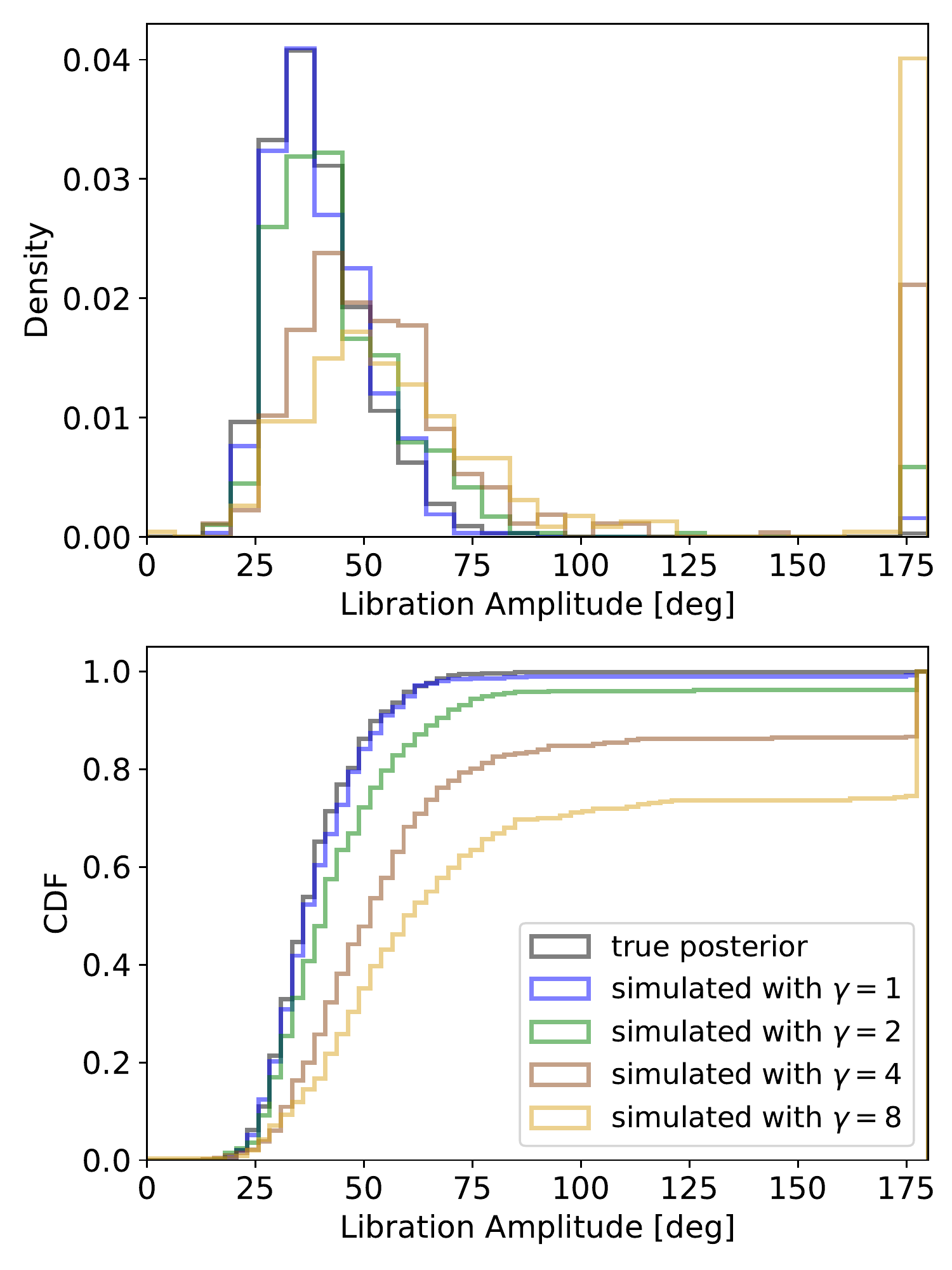}
\caption{Libration amplitude distributions for the critical angle $\phi_1 = 2\lambda_2 - \lambda_1 - \varpi_1$ in the HD 128311 system. The result corresponding to the true posterior distribution from \cite{2015MNRAS.448L..58R} is shown in gray. Additionally, we show the results from four simulated posterior distributions with broadening factors, $\gamma$, indicated in the legend. The top and bottom panels correspond to normalized histograms and cumulative histograms, respectively.} 
\label{fig: libration amplitude distributions}
\end{figure}

We now calculate the distributions of libration amplitudes resulting from both the true and simulated posterior distributions. For each posterior sample, we use the system parameters as initial conditions and run a $1,000$-year $N$-body integration using the REBOUND gravitational dynamics software package \citep{2012A&A...537A.128R} with the ``WHFast'' Wisdom-Holman symplectic integrator \citep{1991AJ....102.1528W, 2015MNRAS.448L..58R}. We calculate the critical angle $\phi_1 = 2\lambda_2 - \lambda_1 - \varpi_1$ and numerically estimate its amplitude using ${A_{\mathrm{lib}} = 0.5(\max{\phi_1} - \min{\phi_1})}$.\footnote{We note that calculating $A_{\mathrm{lib}}$ from the osculating orbital elements in this manner may be another source of overestimation. This is because the osculating elements are affected by high-frequency variations at the synodic period, thus creating a non-zero minimum $A_{\mathrm{lib}}$. This is probably only relevant in systems with very massive planets and tightly spaced MMRs.}

The resulting distributions of libration amplitudes are shown in Figure \ref{fig: libration amplitude distributions}. Here we observe, first, that the distribution resulting from the simulated posterior with $\gamma = 1$ (no broadening) closely resembles that of the true posterior, which provides further confirmation that the multivariate Gaussian approximation is appropriate. Moreover, we observe that the distributions corresponding to the simulated posteriors with $\gamma > 1$ are shifted to progressively larger libration amplitudes as the broadening factor increases. This result offers support of our primary hypothesis. Namely, the libration amplitude distribution does indeed appear to be systematically biased high as a result of noisier data, when simulated in terms of broader posterior distributions.   

\section{Synthetic Data Experiments}
\label{sec: synthetic data experiments}

\begin{figure}
\centering
\epsscale{1.1}
\plotone{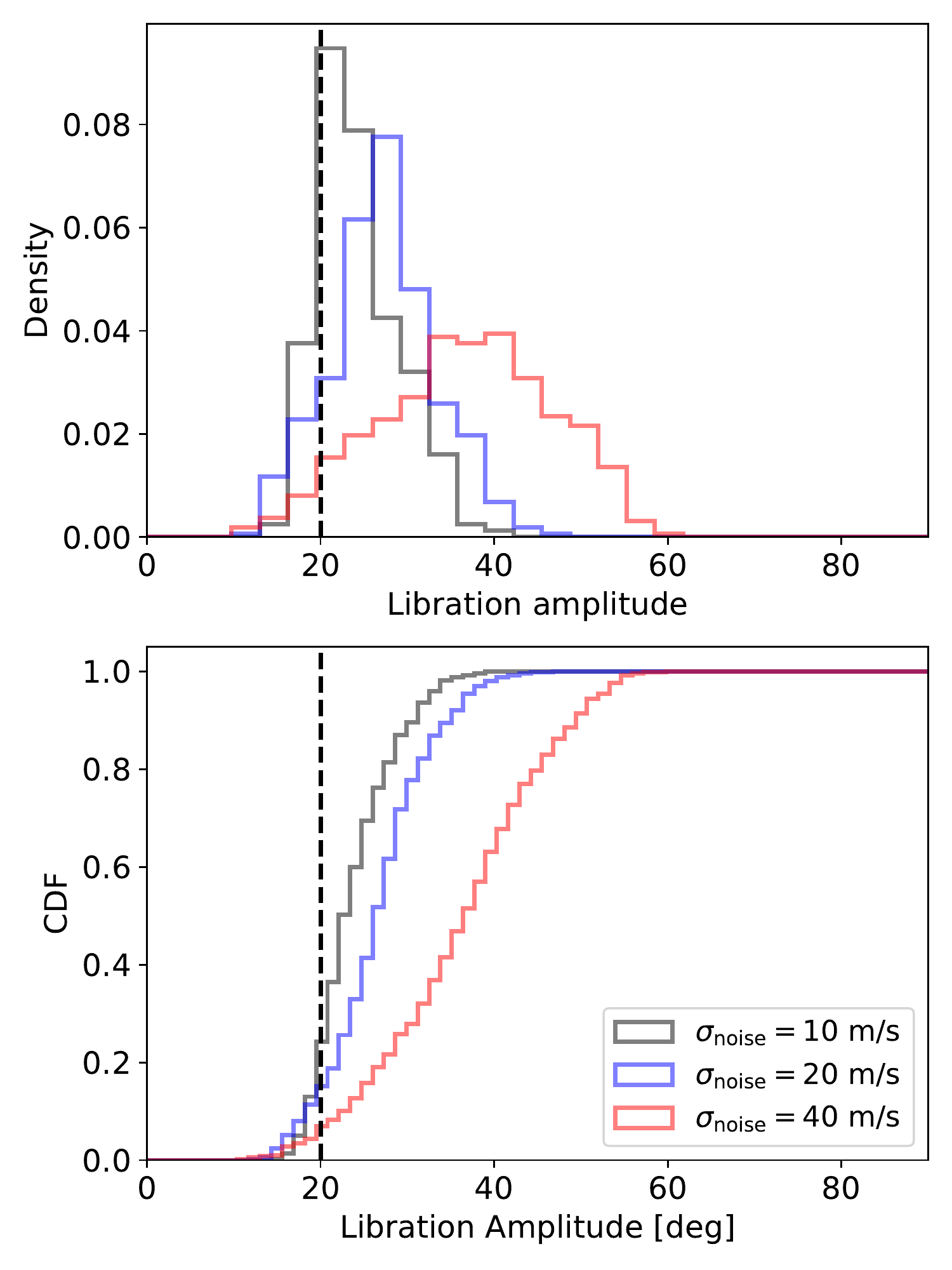}
\caption{Libration amplitude distributions for the critical angle $\phi_1 = 2\lambda_2 - \lambda_1 - \varpi_1$ in a synthetic system closely resembling HD 128311. These results correspond to the posterior distributions resulting from the fits to the synthetic radial velocity data with varying levels of Gaussian noise with standard deviation, $\sigma_{\mathrm{noise}}$. The top and bottom panels correspond to normalized histograms and cumulative histograms, respectively. The dashed black line corresponds to the ``true'' libration amplitude of the synthetic system.}
\label{fig: libration amplitude distributions for synthetic data}
\end{figure}

While the previous exploration of simulated posterior distributions was supportive of our hypothesis, a more thorough examination of the hypothesis would involve systematically varying the signal-to-noise of the data. In this section, we perform dynamical fits to synthetic RV data with various levels of added noise and compare the resulting libration amplitude distributions. 

We use the general parameters of the HD 128311 system. Specifically, we examine the true posterior distribution and extract the parameters of the single sample that we determined to have the lowest libration amplitude, which turns out to be $\sim20^{\circ}$. We use the lowest libration amplitude configuration because we want to explore progressively larger amounts of measurement noise and see how the libration amplitude distribution shifts. We use REBOUND with the IAS15 integrator \citep{2015MNRAS.446.1424R} to generate the synthetic RV measurements with 200 data points randomly spaced over a period of 15 years. We add Gaussian noise with varying standard deviations, $\sigma_{\mathrm{noise}}$. 

Next, we employ the affine-invariant ensemble sampler \texttt{emcee} \citep{2010CAMCS...5...65G, 2013PASP..125..306F} to estimate the posterior distributions of the parameters consistent with the synthetic data. For simplicity, we assume uninformative priors for all parameters. We use 50 walkers and sample the parameters in the same coordinate system as indicated in equation \ref{eq: X_post}. We run the integration for 750 iterations and check for convergence by visual inspection of the log-probability. Finally, we take a random subset of 500 posterior samples from the chains (post burn-in) and use the procedure described in the previous section to calculate the corresponding libration amplitude distributions.

The resulting libration amplitude distributions are shown in Figure \ref{fig: libration amplitude distributions for synthetic data}. Similar to Figure \ref{fig: libration amplitude distributions}, the distributions are shifted to larger libration amplitudes as $\sigma_{\mathrm{noise}}$ increases. Moreover, even the distribution corresponding to the lowest level of noise is still systematically shifted to larger values than $20^{\circ}$, which is the true libration amplitude of the synthetic system. This experiment thus offers a strong confirmation of our hypothesis. That is, any amount of measurement noise will tend to systematically bias the libration amplitude distribution inferred from the posterior distribution, and the degree of the bias increases with the measurement noise.

\section{Discussion}
\label{sec: discussion}

\subsection{Origin of the bias}
\label{sec: origin of the bias}

\begin{figure}
\centering
\epsscale{1.1}
\plotone{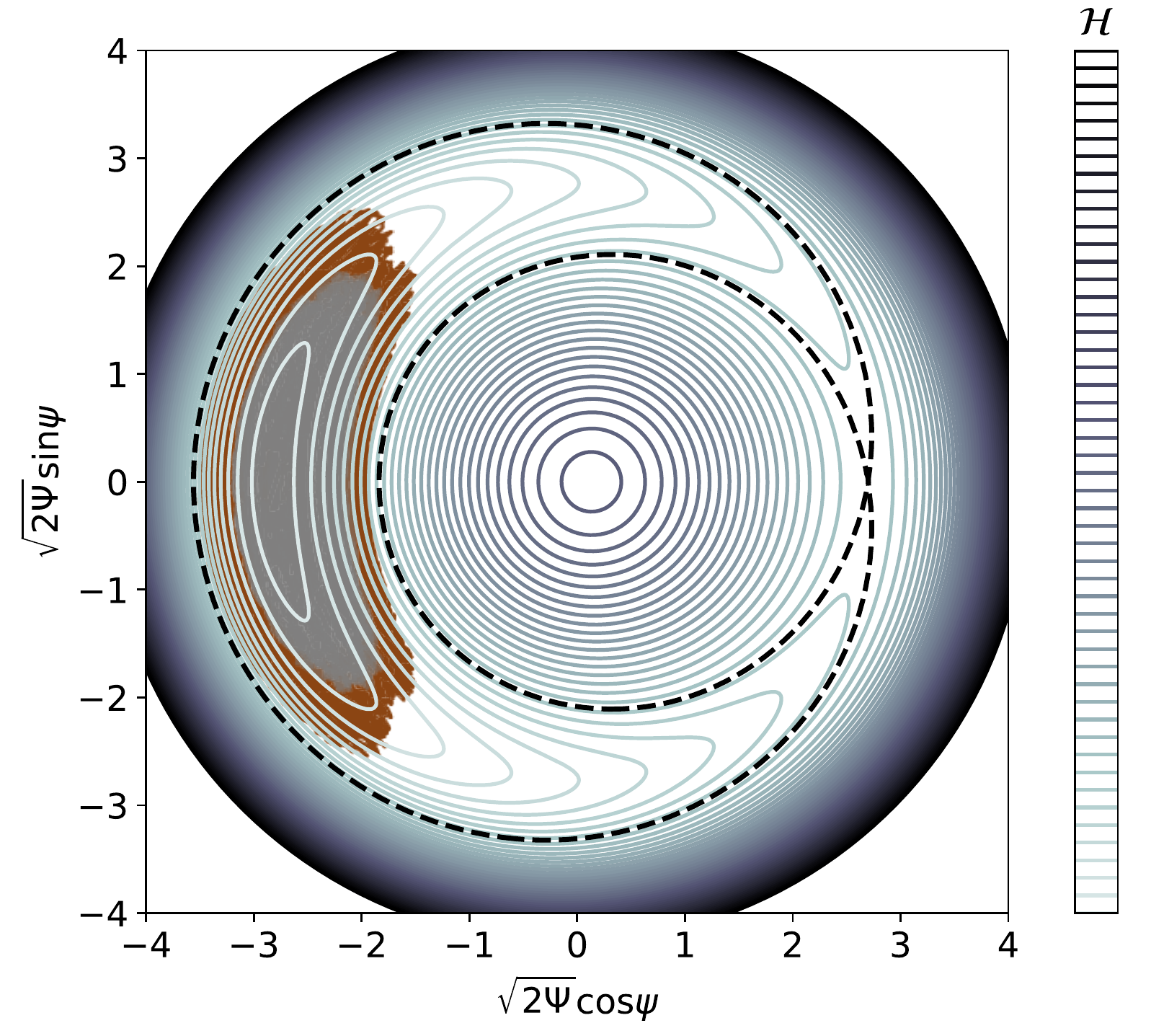}
\caption{Phase space portrait of the approximate first-order resonant motion of HD 128311. The action-angle variables are complicated functions of the planetary orbital elements, but roughly speaking, $\Psi$ is proportional to $\sim e^2$, and $\psi$ is a combination of the critical angles. The level curves correspond to energy levels of the integrable one-degree-of-freedom Hamiltonian from \cite{2013A&A...556A..28B}. The dashed line indicates the separatrix, which encloses the crescent-moon-shaped resonant domain. The gray and brown regions indicate several trajectories derived from integrations of the system with parameters drawn from the true posterior distribution (gray) and the simulated distribution with $\gamma = 2$ (brown). } 
\label{fig: resonant phase space}
\end{figure}

The last two sections confirmed that the bias is indeed real. However, we haven't yet discussed its physical origin. The bias is likely caused by the fact that 
the available phase space volume increases non-uniformly with increasing libration amplitude. That is, if the dynamics are expressed in terms of an integrable one-degree-of-freedom approximation using Hamiltonian perturbation theory \citep[e.g.][]{1983CeMec..30..197H, 2013A&A...556A..28B, 2016ApJ...823...72N}, then the resonant trajectories in phase space are those that are librating inside a finite resonant domain. Only a subset of this domain is available for libration amplitudes below a certain threshold. If we denote $V(A_{\mathrm{lib}})$ as the total phase space volume occupied by resonant trajectories with libration amplitudes $\leq A_{\mathrm{lib}}$, then $dV/dA_{\mathrm{lib}}$ is a positive and increasing function of $A_{\mathrm{lib}}$. Thus, when the posterior distribution of system parameters is broader due to the effects of noise, a uniform sampling of the available resonant phase space will be increasingly skewed to larger libration amplitudes.

Figure \ref{fig: resonant phase space} demonstrates a schematic representation of this. It shows a phase space portrait of the first-order resonant Hamiltonian derived by \cite{2013A&A...556A..28B} and applied to the parameters of the HD 128311 system. We superimpose trajectories resulting from $N$-body integrations of posterior samples, both from the true posterior distribution and the simulated posterior distribution with $\gamma=2$ (Section \ref{sec: exploring the bias}). This illustration indicates that the trajectories of initial system parameters from the broadened posterior extend to a larger region of the resonant domain (and a correspondingly larger range of libration amplitudes) than the trajectories resulting from the true posterior. 

Given a set of initial orbital elements, a single trajectory would ideally fall upon a single level curve. There are several reasons why the gray and brown regions are blurred out and do not follow the topology exactly. First, the analytic approximation assumes small eccentricities, but the eccentricities of the HD 128311 system ($e_1\sim0.3$, $e_2\sim0.12$) are moderate. Second, we are plotting multiple trajectories with different initial conditions, each associated with different libration amplitudes. Finally, the topology of the phase space (as indicated by the level curves) is conserved in time but varies with respect to different sets of initial system parameters. Thus, the level curves in Figure \ref{fig: resonant phase space} can only be thought of as an average representation of the system parameters. Despite these caveats, this schematic illustration helps provide a geometric understanding of how broader posterior distributions of system parameters translate into broader available volumes in the resonant phase space and larger libration amplitudes.

\subsection{Potential remedies}
\label{sec: potential remedies}

\begin{figure}
\centering
\epsscale{1.1}
\plotone{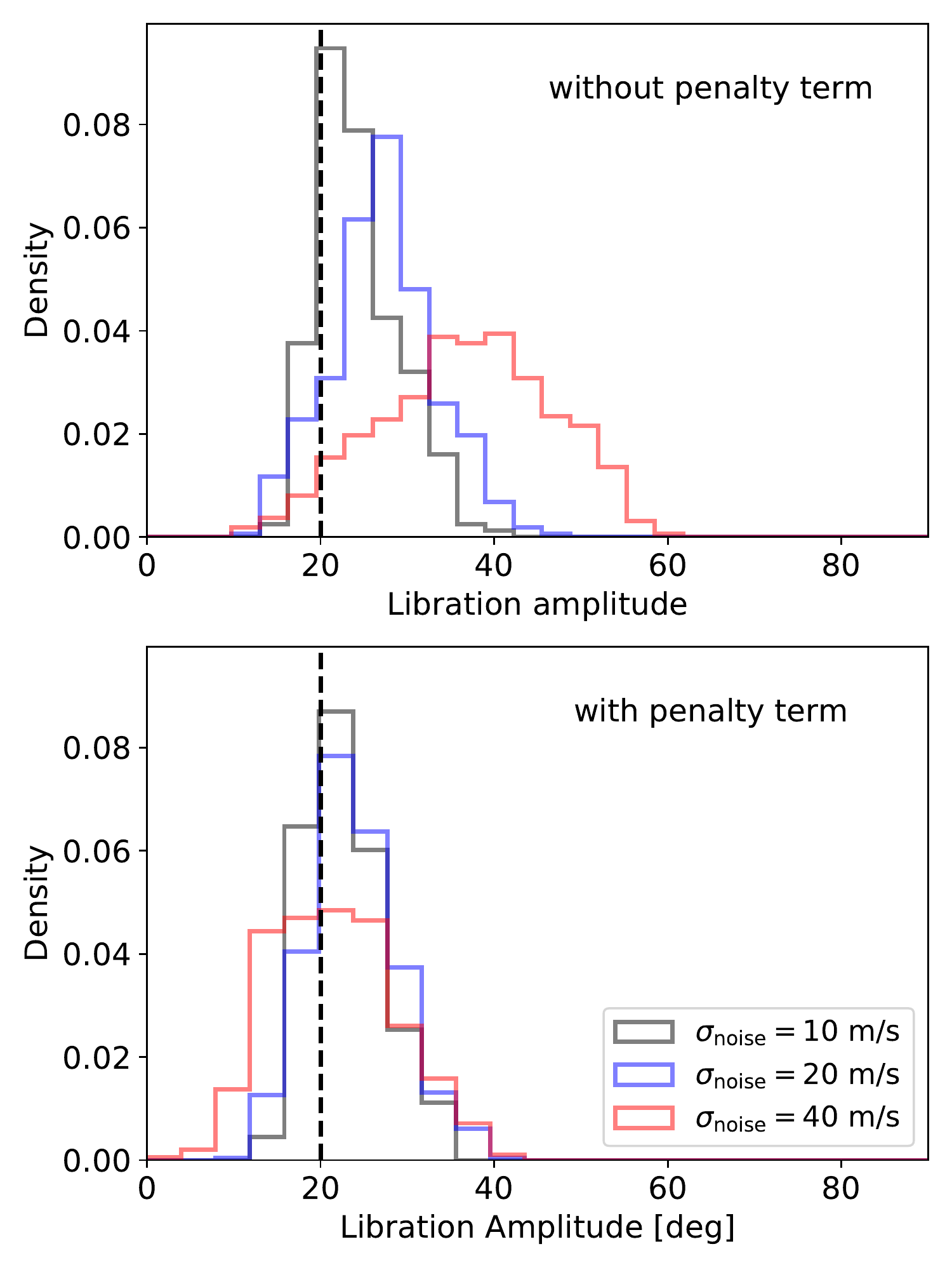}
\caption{Libration amplitude distributions for the critical angle $\phi_1 = 2\lambda_2 - \lambda_1 - \varpi_1$ in a synthetic system closely resembling HD 128311. The top panel is a duplicate of the Figure \ref{fig: libration amplitude distributions for synthetic data} top panel for convenience. The bottom panel corresponds to the posterior distributions resulting from the MCMC sampling with a Gaussian-like penalty term included in the prior. The dashed black line corresponds to the ``true'' libration amplitude of the synthetic system. The medians of the $\sigma_{\mathrm{noise}}=$ 10 m/s, 20 m/s, and 40 m/s distributions are $23^{\circ}$, $27^{\circ}$, and $37^{\circ}$ for the top panel and $22^{\circ}$, $23^{\circ}$, and $21^{\circ}$ for the bottom panel.} 
\label{fig: libration amplitude distributions for synthetic data with prior}
\end{figure}

There are some potential strategies for remedying the bias. One approach would be to set a prior on $A_{\mathrm{lib}}$ to counteract the effect of the bias. The key is to realize that the bias itself is a result of the particular choice of priors. When using the uninformative priors conventionally adopted in RV models, we found in Section \ref{sec: synthetic data experiments} that the marginalized posterior distributions of libration amplitudes are weighted to progressively larger values as a function of increasing measurement noise. The prior on $A_{\mathrm{lib}}$ is implicit in the conventional framework, but it can still be modified. 

For example, when computing the posterior of a proposed MCMC sample, we can include a Gaussian-like penalty term in the prior, such that the prior on $A_{\mathrm{lib}}$ becomes the product of the conventional implicit prior and the penalty term. We test this approach by repeating our synthetic data experiments from Section \ref{sec: synthetic data experiments}, everything being equal except with the inclusion of the penalty term, $\exp[-A_{\mathrm{lib}}^2/(2\sigma^2)]$, which is multiplied by the usual uninformative prior. In order to calculate $A_{\mathrm{lib}}$ for each proposed sample, we perform an additional ``WHFast'' integration with a timestep equal to {40 days}, approximately 8.7\% of $P_1$,\footnote{We note that this timestep is larger than what is generally recommended \citep{2015AJ....150..127W}. We adopted it for computational speed and verified that the resulting $A_{\mathrm{lib}}$ calculations are not significantly affected.}and a duration of 350 years. Performing this extra integration increases the computation time of the posterior by a factor of $\sim3$. As for the $\sigma$ in the penalty term, we find through trial and error that $\sigma\approx0.2$ (when $A_{\mathrm{lib}}$ is in radians) is an appropriate value, in the sense that it yields the desired effect on the libration amplitude distribution, as we will next show. 

Figure \ref{fig: libration amplitude distributions for synthetic data with prior} shows the resulting libration amplitude distributions when the penalty term is used. Compared to the previous case, the distributions agree much better with the true libration amplitude of the synthetic system. The distributions are broadened with increasing $\sigma_{\mathrm{noise}}$, but they do not have the strong systematic shifts seen earlier. Accordingly, this approach is a reasonable ``quick fix'' to approximately counteract the bias. We caution that the optimal value of $\sigma$ in the penalty term can depend on the system at hand, so in terms of an application to observed systems, one should explore a range of different $\sigma$ values and examine the resulting sensitivity of the inferred system parameters to the prior.

A more formal approach to address the bias would be to construct the RV model with specific assumptions about the libration amplitude. For instance, \cite{2020AJ....160..106H} developed an approach for modeling RV data of resonant systems in which the system is assumed to reside in a particular MMR configuration called an ``apsidal corotation resonance'' (ACR). The ACR is the expected outcome of resonant capture under the influence of smooth convergent migration and eccentricity damping, and it has zero libration amplitude. \cite{2020AJ....160..106H} used Bayesian model comparison to assess the evidence in RV data for the ACR model versus a conventional model, and they identified several systems in which the zero-$A_{\mathrm{lib}}$ ACR model is preferred. It is possible that such a model could be extended to include finite libration amplitude configurations as well.

In the case where the conventional approach is to be adopted with no modifications, it is important to be aware of the existence of the bias, particularly when making broader statements on the basis of the inferred libration amplitude. One could explore tests such as varying the size of the dataset going into the fit and seeing how the resulting libration amplitude distributions differ. This could help determine whether or not the distribution is converging. It is also generally appropriate to assume that the true libration amplitude is more closely approximated by values towards the low end of the inferred distribution, as opposed to the mean or median (e.g. Figure \ref{fig: libration amplitude distributions for synthetic data}).

\section{Conclusion}

Exoplanetary systems containing mean-motion resonances offer valuable tests of planet migration and protoplanetary disk properties. However, a key ingredient in an accurate characterization of a resonance is a reliable estimate of the libration amplitude, or the range of oscillations of the critical angle, which indicate how ``deep'' in resonance a system actually is. Motivated by prior work on the GJ 876 system by \cite{2018AJ....155..106M}, here we showed that the reliability of libration amplitude inferences from observations of resonant systems depends sensitively on the data quality and the degree of measurement noise. Specifically, when using conventionally-adopted uninformative priors, progressively larger measurement noise causes the libration amplitude distribution inferred from the posterior distribution of model parameters to be strongly biased towards larger values.

We showed this using two methods of scrutiny of a representative resonant system, HD 128311, which contains two eccentric super-Jupiter-sized planets in a 2:1 MMR with orbital periods equal to $\sim460$ days and $\sim910$ days \citep{2015MNRAS.448L..58R}. The first approach involved a simulated broadening of the posterior distribution of system parameters, mimicking the effects of increasing measurement noise. The second approach involved performing dynamical fits of synthetic data with various levels of noise. In both approaches, the resulting libration amplitude distributions were found to systematically shift to larger values with more noise. Moreover, the synthetic data experiments revealed that even low levels of measurement noise result in inferred libration amplitude distributions that are systematically larger than the ``true'' value.

We highlighted some strategies for mitigating the bias. Specifically, we showed how the simple inclusion of a Gaussian-like penalty term in the prior can avoid the posterior distribution being weighted to large libration amplitudes. Other modifications to the prior are also possible. If the conventional approach is still to be adopted with no modifications, one can examine the extent of the inevitable bias by observing the width of the libration amplitude distribution. (A broader distribution generally indicates a stronger bias.) Another approach is to vary the size of the dataset that is going into the parameter inference and observe whether the libration amplitude distribution is converging or varying strongly with the size of the dataset. In general, an awareness of the existence of the bias will strengthen our ability to characterize resonant systems and to decipher their formation histories.

\section{Acknowledgements}
We are very grateful to the referee, Sam Hadden, for his insightful comments and valuable suggestions, particularly with regard to the content in Sections \ref{sec: origin of the bias} and \ref{sec: potential remedies}. We are also grateful to Hanno Rein for sharing the posterior samples of HD 128311 from \cite{2015MNRAS.448L..58R}. We also thank Eric Ford, Dan Foreman-Mackey, and Greg Laughlin for helpful conversations, as well as Neta Bahcall for her help in facilitating this work through Princeton's Junior Project requirement. S.C.M. was supported by NASA through the NASA Hubble Fellowship grant \#HST-HF2-51465 awarded by the Space Telescope Science Institute, which is operated by the Association of Universities for Research in Astronomy, Inc., for NASA, under contract NAS5-26555.

\bibliographystyle{aasjournal}
\bibliography{main}

\begin{thebibliography}{}
\expandafter\ifx\csname natexlab\endcsname\relax\def\natexlab#1{#1}\fi
\providecommand{\url}[1]{\href{#1}{#1}}
\providecommand{\dodoi}[1]{doi:~\href{http://doi.org/#1}{\nolinkurl{#1}}}
\providecommand{\doeprint}[1]{\href{http://ascl.net/#1}{\nolinkurl{http://ascl.net/#1}}}
\providecommand{\doarXiv}[1]{\href{https://arxiv.org/abs/#1}{\nolinkurl{https://arxiv.org/abs/#1}}}

\bibitem[{{Adams} {et~al.}(2008){Adams}, {Laughlin}, \&
  {Bloch}}]{2008ApJ...683.1117A}
{Adams}, F.~C., {Laughlin}, G., \& {Bloch}, A.~M. 2008, \apj, 683, 1117,
  \dodoi{10.1086/589986}

\bibitem[{{Batygin} {et~al.}(2015){Batygin}, {Deck}, \&
  {Holman}}]{2015AJ....149..167B}
{Batygin}, K., {Deck}, K.~M., \& {Holman}, M.~J. 2015, \aj, 149, 167,
  \dodoi{10.1088/0004-6256/149/5/167}

\bibitem[{{Batygin} \& {Morbidelli}(2013)}]{2013A&A...556A..28B}
{Batygin}, K., \& {Morbidelli}, A. 2013, \aap, 556, A28,
  \dodoi{10.1051/0004-6361/201220907}

\bibitem[{{Butler} {et~al.}(2003){Butler}, {Marcy}, {Vogt}, {Fischer}, {Henry},
  {Laughlin}, \& {Wright}}]{2003ApJ...582..455B}
{Butler}, R.~P., {Marcy}, G.~W., {Vogt}, S.~S., {et~al.} 2003, \apj, 582, 455,
  \dodoi{10.1086/344570}

\bibitem[{{Dawson} {et~al.}(2021){Dawson}, {Huang}, {Brahm}, {Collins},
  {Hobson}, {Jord{\'a}n}, {Dong}, {Korth}, {Trifonov}, {Abe}, {Agabi}, {Bruni},
  {Butler}, {Barbieri}, {Collins}, {Conti}, {Crane}, {Crouzet}, {Dransfield},
  {Evans}, {Espinoza}, {Gan}, {Guillot}, {Henning}, {Lissauer}, {Jensen},
  {Sainte}, {M{\'e}karnia}, {Myers}, {Nandakumar}, {Relles}, {Sarkis},
  {Torres}, {Shectman}, {Schmider}, {Shporer}, {Stockdale}, {Teske}, {Triaud},
  {Wang}, {Ziegler}, {Ricker}, {Vanderspek}, {Latham}, {Seager}, {Winn},
  {Jenkins}, {Bouma}, {Burt}, {Charbonneau}, {Levine}, {McDermott}, {McLean},
  {Rose}, {Vanderburg}, \& {Wohler}}]{2021AJ....161..161D}
{Dawson}, R.~I., {Huang}, C.~X., {Brahm}, R., {et~al.} 2021, \aj, 161, 161,
  \dodoi{10.3847/1538-3881/abd8d0}

\bibitem[{{Delisle}(2017)}]{2017A&A...605A..96D}
{Delisle}, J.~B. 2017, \aap, 605, A96, \dodoi{10.1051/0004-6361/201730857}

\bibitem[{{Dempsey} \& {Nelson}(2018)}]{2018ApJ...867...75D}
{Dempsey}, A.~M., \& {Nelson}, B.~E. 2018, \apj, 867, 75,
  \dodoi{10.3847/1538-4357/aae36c}

\bibitem[{{Fabrycky} {et~al.}(2014){Fabrycky}, {Lissauer}, {Ragozzine}, {Rowe},
  {Steffen}, {Agol}, {Barclay}, {Batalha}, {Borucki}, {Ciardi}, {Ford},
  {Gautier}, {Geary}, {Holman}, {Jenkins}, {Li}, {Morehead}, {Morris},
  {Shporer}, {Smith}, {Still}, \& {Van Cleve}}]{2014ApJ...790..146F}
{Fabrycky}, D.~C., {Lissauer}, J.~J., {Ragozzine}, D., {et~al.} 2014, \apj,
  790, 146, \dodoi{10.1088/0004-637X/790/2/146}

\bibitem[{{Foreman-Mackey} {et~al.}(2013){Foreman-Mackey}, {Hogg}, {Lang}, \&
  {Goodman}}]{2013PASP..125..306F}
{Foreman-Mackey}, D., {Hogg}, D.~W., {Lang}, D., \& {Goodman}, J. 2013, \pasp,
  125, 306, \dodoi{10.1086/670067}

\bibitem[{{Gillon} {et~al.}(2017){Gillon}, {Triaud}, {Demory}, {Jehin}, {Agol},
  {Deck}, {Lederer}, {de Wit}, {Burdanov}, {Ingalls}, {Bolmont}, {Leconte},
  {Raymond}, {Selsis}, {Turbet}, {Barkaoui}, {Burgasser}, {Burleigh}, {Carey},
  {Chaushev}, {Copperwheat}, {Delrez}, {Fernandes}, {Holdsworth}, {Kotze}, {Van
  Grootel}, {Almleaky}, {Benkhaldoun}, {Magain}, \&
  {Queloz}}]{2017Natur.542..456G}
{Gillon}, M., {Triaud}, A. H.~M.~J., {Demory}, B.-O., {et~al.} 2017, \nat, 542,
  456, \dodoi{10.1038/nature21360}

\bibitem[{{Goldreich} \& {Schlichting}(2014)}]{2014AJ....147...32G}
{Goldreich}, P., \& {Schlichting}, H.~E. 2014, \aj, 147, 32,
  \dodoi{10.1088/0004-6256/147/2/32}

\bibitem[{{Goldreich} \& {Tremaine}(1980)}]{1980ApJ...241..425G}
{Goldreich}, P., \& {Tremaine}, S. 1980, \apj, 241, 425, \dodoi{10.1086/158356}

\bibitem[{{Goodman} \& {Weare}(2010)}]{2010CAMCS...5...65G}
{Goodman}, J., \& {Weare}, J. 2010, Communications in Applied Mathematics and
  Computational Science, 5, 65, \dodoi{10.2140/camcos.2010.5.65}

\bibitem[{{Hadden} \& {Payne}(2020)}]{2020AJ....160..106H}
{Hadden}, S., \& {Payne}, M.~J. 2020, \aj, 160, 106,
  \dodoi{10.3847/1538-3881/aba751}

\bibitem[{{Henrard} \& {Lemaitre}(1983)}]{1983CeMec..30..197H}
{Henrard}, J., \& {Lemaitre}, A. 1983, Celestial Mechanics, 30, 197,
  \dodoi{10.1007/BF01234306}

\bibitem[{{Hogg} {et~al.}(2010){Hogg}, {Myers}, \&
  {Bovy}}]{2010ApJ...725.2166H}
{Hogg}, D.~W., {Myers}, A.~D., \& {Bovy}, J. 2010, \apj, 725, 2166,
  \dodoi{10.1088/0004-637X/725/2/2166}

\bibitem[{{H{\"u}hn} {et~al.}(2021){H{\"u}hn}, {Pichierri}, {Bitsch}, \&
  {Batygin}}]{2021A&A...656A.115H}
{H{\"u}hn}, L.~A., {Pichierri}, G., {Bitsch}, B., \& {Batygin}, K. 2021, \aap,
  656, A115, \dodoi{10.1051/0004-6361/202142176}

\bibitem[{{Laughlin} {et~al.}(2005){Laughlin}, {Butler}, {Fischer}, {Marcy},
  {Vogt}, \& {Wolf}}]{2005ApJ...622.1182L}
{Laughlin}, G., {Butler}, R.~P., {Fischer}, D.~A., {et~al.} 2005, \apj, 622,
  1182, \dodoi{10.1086/424686}

\bibitem[{{Lucy} \& {Sweeney}(1971)}]{1971AJ.....76..544L}
{Lucy}, L.~B., \& {Sweeney}, M.~A. 1971, \aj, 76, 544, \dodoi{10.1086/111159}

\bibitem[{{Marcy} {et~al.}(2001){Marcy}, {Butler}, {Fischer}, {Vogt},
  {Lissauer}, \& {Rivera}}]{2001ApJ...556..296M}
{Marcy}, G.~W., {Butler}, R.~P., {Fischer}, D., {et~al.} 2001, \apj, 556, 296,
  \dodoi{10.1086/321552}

\bibitem[{{McArthur} {et~al.}(2014){McArthur}, {Benedict}, {Henry}, {Hatzes},
  {Cochran}, {Harrison}, {Johns-Krull}, \& {Nelan}}]{2014ApJ...795...41M}
{McArthur}, B.~E., {Benedict}, G.~F., {Henry}, G.~W., {et~al.} 2014, \apj, 795,
  41, \dodoi{10.1088/0004-637X/795/1/41}

\bibitem[{{Millholland} \& {Laughlin}(2019)}]{2019NatAs...3..424M}
{Millholland}, S., \& {Laughlin}, G. 2019, Nature Astronomy, 3, 424,
  \dodoi{10.1038/s41550-019-0701-7}

\bibitem[{{Millholland} {et~al.}(2018){Millholland}, {Laughlin}, {Teske},
  {Butler}, {Burt}, {Holden}, {Vogt}, {Crane}, {Shectman}, \&
  {Thompson}}]{2018AJ....155..106M}
{Millholland}, S., {Laughlin}, G., {Teske}, J., {et~al.} 2018, \aj, 155, 106,
  \dodoi{10.3847/1538-3881/aaa894}

\bibitem[{{Mills} {et~al.}(2016){Mills}, {Fabrycky}, {Migaszewski}, {Ford},
  {Petigura}, \& {Isaacson}}]{2016Natur.533..509M}
{Mills}, S.~M., {Fabrycky}, D.~C., {Migaszewski}, C., {et~al.} 2016, \nat, 533,
  509, \dodoi{10.1038/nature17445}

\bibitem[{{Murray} \& {Dermott}(1999)}]{1999ssd..book.....M}
{Murray}, C.~D., \& {Dermott}, S.~F. 1999, {Solar system dynamics}

\bibitem[{{Nelson} {et~al.}(2016){Nelson}, {Robertson}, {Payne}, {Pritchard},
  {Deck}, {Ford}, {Wright}, \& {Isaacson}}]{2016MNRAS.455.2484N}
{Nelson}, B.~E., {Robertson}, P.~M., {Payne}, M.~J., {et~al.} 2016, \mnras,
  455, 2484, \dodoi{10.1093/mnras/stv2367}

\bibitem[{{Nesvorn{\'y}} {et~al.}(2022){Nesvorn{\'y}}, {Chrenko}, \&
  {Flock}}]{2022ApJ...925...38N}
{Nesvorn{\'y}}, D., {Chrenko}, O., \& {Flock}, M. 2022, \apj, 925, 38,
  \dodoi{10.3847/1538-4357/ac36cd}

\bibitem[{{Nesvorn{\'y}} \& {Vokrouhlick{\'y}}(2016)}]{2016ApJ...823...72N}
{Nesvorn{\'y}}, D., \& {Vokrouhlick{\'y}}, D. 2016, \apj, 823, 72,
  \dodoi{10.3847/0004-637X/823/2/72}

\bibitem[{{Peale}(1976)}]{1976ARA&A..14..215P}
{Peale}, S.~J. 1976, \araa, 14, 215,
  \dodoi{10.1146/annurev.aa.14.090176.001243}

\bibitem[{{Rein}(2015)}]{2015MNRAS.448L..58R}
{Rein}, H. 2015, \mnras, 448, L58, \dodoi{10.1093/mnrasl/slu202}

\bibitem[{{Rein} \& {Liu}(2012)}]{2012A&A...537A.128R}
{Rein}, H., \& {Liu}, S.~F. 2012, \aap, 537, A128,
  \dodoi{10.1051/0004-6361/201118085}

\bibitem[{{Rein} \& {Papaloizou}(2009)}]{2009A&A...497..595R}
{Rein}, H., \& {Papaloizou}, J.~C.~B. 2009, \aap, 497, 595,
  \dodoi{10.1051/0004-6361/200811330}

\bibitem[{{Rein} \& {Spiegel}(2015)}]{2015MNRAS.446.1424R}
{Rein}, H., \& {Spiegel}, D.~S. 2015, \mnras, 446, 1424,
  \dodoi{10.1093/mnras/stu2164}

\bibitem[{{Rivera} {et~al.}(2010){Rivera}, {Laughlin}, {Butler}, {Vogt},
  {Haghighipour}, \& {Meschiari}}]{2010ApJ...719..890R}
{Rivera}, E.~J., {Laughlin}, G., {Butler}, R.~P., {et~al.} 2010, \apj, 719,
  890, \dodoi{10.1088/0004-637X/719/1/890}

\bibitem[{{Rivera} {et~al.}(2005){Rivera}, {Lissauer}, {Butler}, {Marcy},
  {Vogt}, {Fischer}, {Brown}, {Laughlin}, \& {Henry}}]{2005ApJ...634..625R}
{Rivera}, E.~J., {Lissauer}, J.~J., {Butler}, R.~P., {et~al.} 2005, \apj, 634,
  625, \dodoi{10.1086/491669}

\bibitem[{{Shen} \& {Turner}(2008)}]{2008ApJ...685..553S}
{Shen}, Y., \& {Turner}, E.~L. 2008, \apj, 685, 553, \dodoi{10.1086/590548}

\bibitem[{{Silburt} \& {Rein}(2017)}]{2017MNRAS.469.4613S}
{Silburt}, A., \& {Rein}, H. 2017, \mnras, 469, 4613,
  \dodoi{10.1093/mnras/stx1193}

\bibitem[{{Terquem} \& {Papaloizou}(2007)}]{2007ApJ...654.1110T}
{Terquem}, C., \& {Papaloizou}, J. C.~B. 2007, \apj, 654, 1110,
  \dodoi{10.1086/509497}

\bibitem[{{Vogt} {et~al.}(2005){Vogt}, {Butler}, {Marcy}, {Fischer}, {Henry},
  {Laughlin}, {Wright}, \& {Johnson}}]{2005ApJ...632..638V}
{Vogt}, S.~S., {Butler}, R.~P., {Marcy}, G.~W., {et~al.} 2005, \apj, 632, 638,
  \dodoi{10.1086/432901}

\bibitem[{{Wisdom}(2015)}]{2015AJ....150..127W}
{Wisdom}, J. 2015, \aj, 150, 127, \dodoi{10.1088/0004-6256/150/4/127}

\bibitem[{{Wisdom} \& {Holman}(1991)}]{1991AJ....102.1528W}
{Wisdom}, J., \& {Holman}, M. 1991, \aj, 102, 1528, \dodoi{10.1086/115978}

\bibitem[{{Wright} {et~al.}(2011){Wright}, {Veras}, {Ford}, {Johnson}, {Marcy},
  {Howard}, {Isaacson}, {Fischer}, {Spronck}, {Anderson}, \&
  {Valenti}}]{2011ApJ...730...93W}
{Wright}, J.~T., {Veras}, D., {Ford}, E.~B., {et~al.} 2011, \apj, 730, 93,
  \dodoi{10.1088/0004-637X/730/2/93}

\end{thebibliography}

\end{document}